\documentclass[aps,prl,twocolumn,superscriptaddress,floatfix,10pt]{revtex4-1}
\usepackage{amssymb,amsmath}
\usepackage{graphicx}
\usepackage{float}
\usepackage{bm}
\usepackage{multirow}
\usepackage{dcolumn}
\usepackage{tabularx}
\usepackage{longtable}
\usepackage[dvipsnames,usenames]{color}
\usepackage[latin1]{inputenc} 
\usepackage{hyperref}

\newlength{\dbarheight}

\begin{document}

\title{Origin of the orbital and spin orderings in rare-earth titanates}

\author{Julien Varignon} \affiliation{Unit\'e Mixte de Physique, CNRS,
  Thales, Universit\'e Paris Sud, Universit\'e Paris-Saclay, 91767,
  Palaiseau, France} \email{julien.varignon@cnrs-thales.fr}

\author{Mathieu N. Grisolia} \affiliation{Unit\'e Mixte de Physique,
  CNRS, Thales, Universit\'e Paris Sud, Universit\'e Paris-Saclay,
  91767, Palaiseau, France}

\author{Daniele Preziosi} \affiliation{Unit\'e Mixte de Physique,
  CNRS, Thales, Universit\'e Paris Sud, Universit\'e Paris-Saclay,
  91767, Palaiseau, France}

\author{Philippe Ghosez} \affiliation{Theoretical Materials Physics,
  Q-MAT, CESAM, Universit\'e de Li\`ege (B5), B-4000 Li\`ege, Belgium}

\author{Manuel Bibes} \affiliation{Unit\'e Mixte de Physique, CNRS,
  Thales, Universit\'e Paris Sud, Universit\'e Paris-Saclay, 91767,
  Palaiseau, France}

\date{\today}
\begin{abstract}
Rare-earth titanates RTiO$_3$ are Mott insulators displaying a rich
physical behavior, featuring most notably orbital and spin orders in
their ground state. The origin of their ferromagnetic to
antiferromagnetic transition as a function of the size of the
rare-earth however remains debated. Here we show on the basis of
symmetry analysis and {\em first-principles} calculations that
although rare-earth titanates are nominally Jahn-Teller active, the
Jahn-Teller distortion is negligible and irrelevant for the
description of the ground state properties. At the same time, we
demonstrate that the combination of two antipolar motions produces an
effective Jahn-Teller-like motion which is the key of the varying
spin-orbital orders appearing in titanates. Thus, titanates are
prototypical examples illustrating how a subtle interplay between
several lattice distortions commonly appearing in perovskites can
produce orbital orderings and insulating phases irrespective of proper
Jahn-Teller motions.
\end{abstract}
\maketitle

ABO$_3$ oxide perovskites, with partly filled $d$ states on the B
site, exhibit a rich physical behavior originating from the intimate
coupling between structural, electronic (charge and orbital) and
magnetic degrees of freedom~\cite{Zubko2011interface}. Typical
examples are the rare-earth vanadates R$^{3+}$V$^{3+}$O$_3$
($3d^2$-$t_{2g}^2$ electronic configuration on V$^{3+}$ ions) that
exhibit two different spin and orbital orders yielding distinct
symmetries for the ground state at low
temperature~\cite{Miyasaka-diagAVO3,Vanadates-Julien}. With the
electronic degeneracy of Ti$^{3+}$ -- 3$d^1$-$t_{2g}^1$ configuration
--, rare-earth titanates R$^{3+}$Ti$^{3+}$O$_3$ are often expected to
be another text book example of such a subtle interplay between
orbital and spin orders.

Rare-earth titanates are Mott insulators, which according to their
small tolerance factor, adopt a common orthorhombic $Pbnm$ structure
characterized by large oxygen cage
rotations~\cite{greedan1985rare,LaTiO3-structure-10K,magnetoeles-titanates},
{\em i.e } a$^-$a$^-$c$^+$ antiferrodistortive motions in Glazer's
notations~\cite{Glazer}. They also all undergo a magnetic phase
transition to either a ferromagnetic (FM) ordering for small R=
Lu-Gd+Y or a G-type antiferromagnetic (G-AFM) ordering for large R=
Sm-La~\cite{katsufuji1997transport,magnetoeles-titanates,NJPhys-Mochizuki-RTO}.

The nature of the very peculiar FM to G-AFM transition as a function
of the rare-earth size is however puzzling and
controversial~\cite{NJPhys-Mochizuki-RTO}. On one hand, Ti$^{3+}$ is
nominally a Jahn-Teller (JT) active ion and the JT distortion is
commonly proposed as a key ingredient to explain the
transition~\cite{mizokawa1999interplay,LTO-nondegenerate}. However,
while such a distortion could be compatible with the ferromagnetic
phase~\cite{mochizuki2000magnetic,mochizuki2001origin}, it cannot
provide a satisfying explanation for the purely antiferromagnetic
phase~\cite{mizokawa1999interplay}. On the other hand, some other
works have proposed that the JT distortion is neither responsible for
the insulating phase of these materials nor for the observed orbital
orders~\cite{Pavarini-PRL-RTO,Pavarini-HowChemistry}.

Instead, Mochizuki {\em et al} have suggested that specific
orbital-orderings for the FM and AFM phases are triggered by the
crystal field produced by the
rare-earth~\cite{mochizuki2001origin,Mochizuki-PRL-RTO}, with a
potential competition with the JT
distortion~\cite{LTO-nondegenerate}. This latter model ultimately
results in combinations of the three $t_{2g}$
orbitals~\cite{mochizuki2001origin,akimitsu2001direct,Mochizuki-PRL-RTO,NJPhys-Mochizuki-RTO}
and now appears as a generic mechanism to yield the coupled
spin-orbital orders in the ground state of 3$d^1$
systems~\cite{Pavarini-PRL-RTO}. However, clear theoretical evidence
of the individual role of each lattice distortions, including the
ubiquitous oxygen cage rotations and/or rare-earths motions, is still
missing.

In this manuscript, we revisit the nature of the orbital and spin
orders in the ground state of rare-earth titanates on the basis of
symmetry mode analysis and {\em first-principles} calculations. While
the JT distortion appears rather negligible, we show that the
combination of two specific antipolar distortions involving the
rare-earth produces an effective JT motion tuning the spin-orbital
properties of the low temperature phase.

\begin{figure}
\begin{center}
\resizebox{7.0cm}{!}{\includegraphics{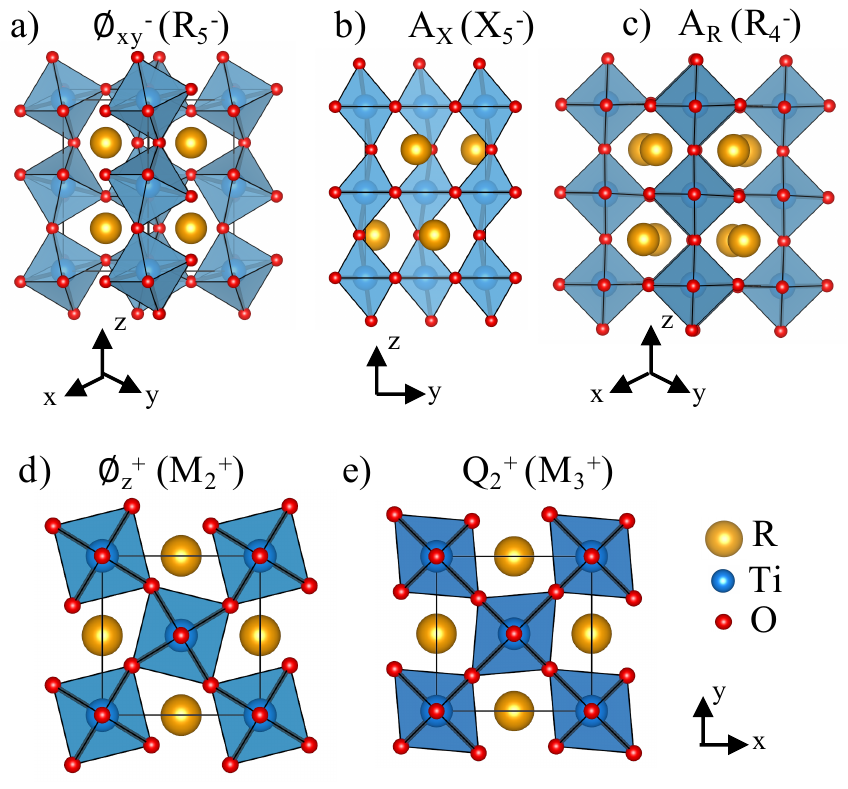}}
\end{center}
\caption{Sketches of the lattice distortions appearing in rare-earth
  titanates ground state.  a) antiphase $\Phi_{xy}^-$ (R$_5^-$ irreps)
  oxygen cage rotation; b) antipolar $A_X$ (irreps X$_5^-$) motion. c)
  antipolar $A_R$ (R$_4^-$ irreps) motion; d) in phase $\Phi_z^+$
  (M$_2^{+}$ irreps) oxygen cage rotation. e) $Q_2^+$ (M$_3^+$ irreps)
  Jahn-Teller motion. Note that amplitudes of distortions have been
  amplified on the sketches and are not representative of their
  magnitude in the ground state structure. }
\label{f:dist}
\end{figure}

\begin{table}[h!]
\begin{center}
\begin{tabular}{ccccccc}
\hline \hline & & Y~\cite{magnetoeles-titanates} &
Gd~\cite{magnetoeles-titanates}& Sm~\cite{magnetoeles-titanates}&
Nd~\cite{magnetoeles-titanates} &La~\cite{LTO-nondegenerate} \\ 
mode & (t. factor) & (0.831) & (0.890) & (0.898) & (0.908) & (0.927)
\\ 
\hline \\ 

\multirow{2}{*}{$\Phi_{xy}^-$ (R$_5^-$)} & exp. & 1.83 & 1.70 & 1.61 &
1.62 & 1.32 \\ & calc. & 1.95 & - &-& -& 1.44 \\
\\ \multirow{2}{*}{$\Phi_z^+$ (M$_2^+$)} & exp. & 1.30& 1.24&
1.17&1.18 &0.95 \\ & calc. & 1.34 & - & - & - & 1.04 \\ \\
\multirow{2}{*}{$A_X$ (X$_ 5^-$)} & exp. & 0.94& 0.86& 0.77&0.71&0.56
\\ & calc. & 0.99 & - & - & - & 0.66 \\ \\
\multirow{2}{*}{$Q_2^+$ (M$_ 3^+$)} & exp. & 0.01 & 0.01&
0.01&0.02&0.04 \\ & calc. & 0.02 & - & - & - & 0.05 \\ \\
\multirow{2}{*}{$A_R$ (R$_4^-$)} & exp. & 0.25& 0.21& 0.17 &0.14 &
0.09 \\ & calc. & 0.29 & - & - & - & 0.11 \\ \hline
\end{tabular}
\end{center}
\caption{Amplitude of distortions (in \AA) of some available
  experimental rare-earth titanates structures. The values for our
  optimized structures (0 K) are also reported.  Experimental
  structures are taken from reference~\cite{LTO-nondegenerate} at 8~K
  for LaTiO$_3$, and from reference~\cite{magnetoeles-titanates} at
  290 K, 100 K, 290 K and 2 K for NdTiO$_3$, SmTiO$_3$, GdTiO$_3$ and
  YTiO$_3$ respectively. The Goldschmidt tolerance factor is given in
  parenthesis and is extracted using tolerance factor calculator
  from~\onlinecite{NicoleBenedek}.}
\label{t:sym-exp}
\end{table}

We first performed a symmetry-adapted mode analysis (with
AMPLIMODES~\cite{Amplimodes1,Amplimodes2}) of some available
experimental data in order to quantify the amplitude of distinct
lattice distortions appearing in titanates. The results are summarized
in Table~\ref{t:sym-exp}. As expected, all titanates develop strong
antiferrodistortive motions -- anti-phase $\Phi_{xy}^-$ (R$_5^-$
irreps) and in-phase $\Phi_z^+$ (M$_2^+$ irreps) motions corresponding
to a$^-$a$^-$c$^0$ and a$^0$a$^0$c$^+$ rotations respectively (see
Figures~\ref{f:dist}.a and d) -- whose strengths are governed by
steric effects. They also exhibit strong $A_X$ (X$_5^-$ irreps) and
$A_R$ (R$_4^-$ irreps) distortions, involving antipolar motions of
rare-earth and/or coplanar oxygens in the (ab)-plane as sketched in
Figures~\ref{f:dist}.b and c. These two modes also seem to be governed
by steric effects, with a softening of their magnitude with increasing
the rare-earth ionic radius albeit the $R_4^-$ mode decreases more
abruptly. Additionally, they also develop a Jahn-Teller (JT)
distortion involving equatorial oxygen motions -- two anions move
inward, two outward -- while apical oxygens are fixed (see
Figure~\ref{f:dist}.e). This motion being in phase between consecutive
planes along the $c$ axis, we label this JT mode as $Q_2^+$ (M$_3^+$
irreps) following Goodenough notation~\cite{JTnotation}. Surprisingly,
this JT distortion is found very weak for all titanates, although it
monotonously increases when going from Y to La. JT distortions are
well known to be smaller for $t_{2g}$ electrons than for $e_g$
electrons but they remain here one order of magnitude smaller than
amplitudes appearing in the ground state of rare-earth vanadates
(V$^{3+}$-$t_{2g}^2$ electronic
degeneracy)~\cite{Vanadates-Julien}. This analysis of experimental
structures therefore provides strong support to the relatively small
contribution of the JT distortion in titanates suggested by former
studies~\cite{Pavarini-HowChemistry}.

In order to gain microscopic insights on the relationship between
these distortions and spin-orbital orders, we performed {\em
  first-principles} calculations using Density Functional Theory (DFT)
with the Vienna Ab-initio Simulation Package~\cite{VASP1,VASP2}. We
used the PBE functional revised for solids~\cite{PBEsol} in
combination with effective Hubbard U$_{\text{eff}}$
corrections~\cite{LDAU-Duda} of 2.5~eV on Ti $3d$ levels and of 1 eV
on the rare-earth $4f$ levels in order to account for the electronic
correlations (see supplementary materials~\cite{SupplementaryMaterial}
for a detailed discussion on the choice of these parameters). We used
the Projector Augmented Waves (PAW) pseudopotentials~\cite{PAW} with
the following valence electron configurations: $4s^23d^2$ (Ti),
$2s^22p^4$ (O), $4s^2$ $4p^65s^25d^14f^0$ (Y), $5p^66s^25d1^14f^0$
(La). An energy cut-off of 500 eV was used and we relaxed geometries
until forces are lower than 1 meV/\AA. A 6$\times$6$\times$4 k-point
grid was used to sample the Brillouin zone unless stated otherwise. We
explored four different magnetic orderings during the calculations:
ferromagnetic (FM), as well as A, C and G-type antiferromagnetic
solutions with spins treated only at the collinear level. We focus in
this study on YTiO$_3$ and LaTiO$_3$ appearing as model systems to
understand the Ti $3d$ electronic structure since they do not possess
$4f$ electrons.

Geometry optimizations for these two compounds yield a $Pbnm$ ground
state associated with a FM and a G-type AFM solution for YTiO$_3$ and
LaTiO$_3$ respectively, consistently with experiments. While the
stability of the FM solution of YTiO$_3$ is rather large ($\Delta$E =
-18.5 meV/f.u. between the FM and G-AFM solutions), the stability of
the G-AFM over the FM solution in LaTiO$_3$ is small ($\Delta$E = -3.4
meV/f.u.) likely underlying a weakly stable AFM solution. The
extracted amplitude of distortions of our ground states are reported
in Table~\ref{t:sym-exp} and are compatible with experimental reports,
therefore validating our optimizations (lattice parameters and atomic
positions are given in supplementary material).

\begin{figure}
\begin{center}
\resizebox{8.6cm}{!}{\includegraphics{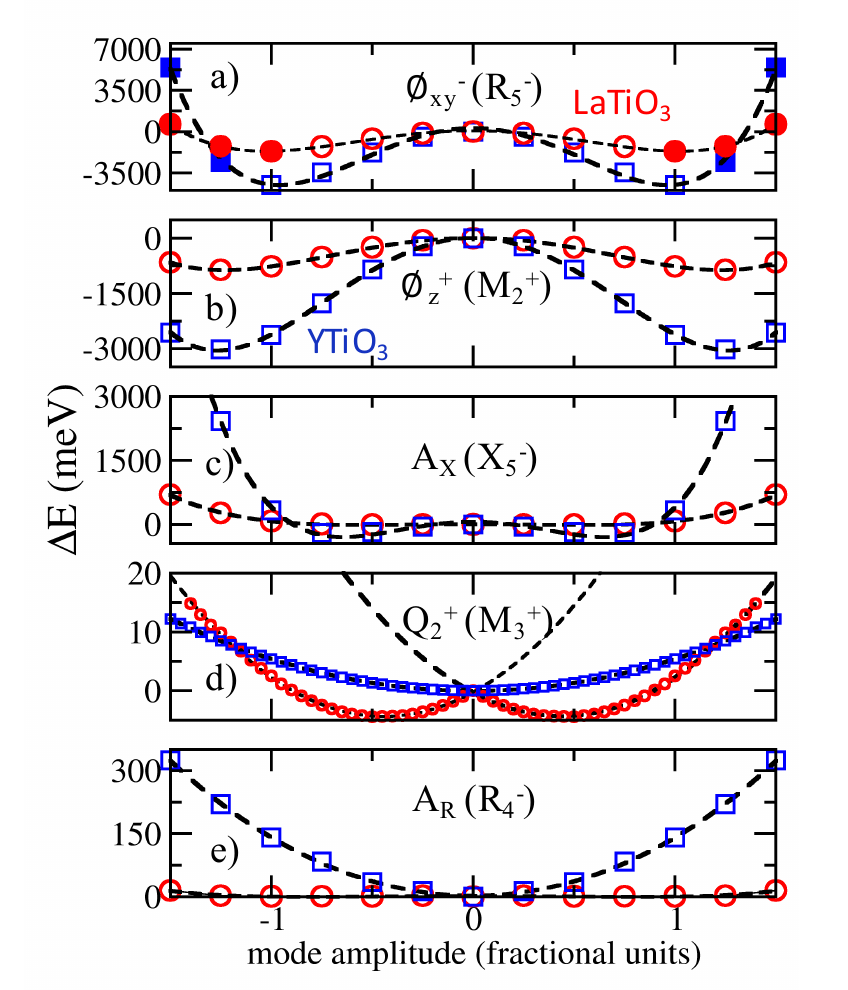}}
\end{center}
\caption{Energy potentials with respect to the amplitude of distortion
  (in fractional units) of the different modes appearing in the ground
  state of YTiO$_3$ (blue squares) and LaTiO$_3$ (red circles), 1.00
  representing the actual distortion appearing in the ground state of
  each material. Filled (unfilled) symbols represent insulating
  (metallic) solutions.  Calculations have been performed in a
  pseudocubic unit cell having the volume of the ground state
  structure. a) the $\Phi_{xy}^-$ mode (a$^-$a$^-$c$^0$ oxygen cage
  rotation). b) the $\Phi_z^+$ mode (a$^0$a$^0$c$^+$ oxygen cage
  rotation). c) the $A_X$ antipolar mode. d) the $Q_2^+$ mode
  corresponding to a Jahn-Teller distortion. e) the $A_R$ antipolar
  mode. We emphasize that the kpoint mesh is increased to $12\times
  12\times 8$ in order to increase both accuracy and convergence of
  the energies.}
\label{f:potentiels}
\end{figure}

We then explored the origin of the orthorhombic structure by studying
the energy potentials of the different lattice distortions. To that
end, we have frozen some lattice motions in a hypothetical
high-symmetry cubic structure ($Pm\bar{3}m$) having the volume of the
ground state structure. Note that the k-point mesh is increased to
$12\times 12\times 8$ in order to enhance both accuracy and
convergence of the wavefunction. Figure~\ref{f:potentiels} reports the
energy landscapes for YTiO$_3$ (blue squares) and LaTiO$_3$ (red
circles) using a FM configuration~\cite{NotePotentiels} (energy
potentials using a G-type AFM configuration are reported in the
supplementary material). As expected, the two oxygen cage rotations
($\Phi_{xy}^-$ and $\Phi_z^+$ modes) present double well potentials
associated with strong energy gains and produce the orthorhombic
$Pbnm$ symmetry. The antipolar $A_X$ mode also develops a double well
potential for YTiO$_3$ and LaTiO$_3$. The energy gain is larger for
YTiO$_3$ albeit one order of magnitude smaller than that of the oxygen
cage rotations. For both compounds, the antipolar $A_R$ mode is always
associated with a single well energy potential, indicating that this
mode is not intrinsically unstable and the driving force of the ground
state.  Importantly, the JT distortion behaves differently for the two
compounds: we do observe single well potentials for YTiO$_3$ and
LaTiO$_3$ but the minimum is shifted to non zero amplitude of the JT
mode for LaTiO$_3$. It is worth noticing that we observe two distinct
wells for LaTiO$_3$ depending on the sign of the lattice
distortion. Therefore, it seems that in LaTiO$_3$, the JT distortion
is able to produce an energy gain by favoring an orbital
polarization. As inferred by the different behavior obtained for the
two compounds, the ability of the JT distortion to produce energy
gains seems to highly rely on the tolerance factor or the unit cell
volume. We can check this hypothesis in our calculations by computing
the energy potentials of YTiO$_3$ and LaTiO$_3$ by using the volume of
the other compound (results are presented in the supplementary
material). When using YTiO$_3$ volume, the JT potential for LaTiO$_3$
becomes a single well potential whose energy minimum is located at
zero amplitude of the JT mode. On the other hand, under a volume
expansion, the minimum of the JT potential of YTiO$_3$ is shifted to
non zero amplitude similarly to LaTiO$_3$ ground
state~\cite{NoteInstabilite}.  This striking result is in close
agreement with the experimental observation of a strengthening of the
amplitude associated with the JT distortion with increasing the
tolerance factor (see table~\ref{t:sym-exp}). Finally, it is worth to
emphasize that only large $\Phi_{xy}^-$ rotation either in the FM or
G-AFM spin ordering for YTiO$_3$ and LaTiO$_3$ -- and large $A_X$
antipolar modes in the G-AFM configuration for YTiO$_3$ (see
supplementary material)--, are able to open a band gap.

Being not necessarily intrinsically unstable, the presence of the JT
and antipolar $A_R$ modes originates from the symmetry allowed terms
in the free energy expansion $\mathcal{F}$ around a $Pm\bar{3}m$ cubic
symmetry. Among all possible terms, $\mathcal{F}$ exhibits several
trilinear couplings:
\begin{eqnarray}
\mathcal{F} & \propto & a\cdot \Phi_{xy}^- \cdot \Phi_z^+ \cdot A_X +
b\cdot \Phi_{xy}^- \cdot A_X \cdot Q_2^+ \nonumber \\ & +& c\cdot
\Phi_z^+ \cdot A_X \cdot A_R + d\cdot A_X \cdot A_R \cdot Q_2^+
\label{e:freeenergy}
\end{eqnarray}
According to the first term, the condensation of the two rotations
($\Phi_{xy}^-$ and $\Phi_z^+$ modes) automatically brings the
antipolar $A_X$ motion in the system in order to lower the energy,
irrespective of its stability/instability. Subsequently, the second
term of equation~\ref{e:freeenergy} will force the appearance of the
JT distortion in any case. The latter therefore has an improper
origin, a mechanism already discussed in some other
systems~\cite{Vanadates-Julien,JT-PRL-Julien,Titanates-Nick} and that
explains its small amplitude for titanates with low tolerance
factor. Finally, according to the third term in
Eq.~\ref{e:freeenergy}, the $A_R$ antipolar mode also appears as a
consequence of the $\Phi_z^+$ ($a^0a^0c^+$) oxygen cage rotation and
the $A_X$ antipolar motion.

These trilinear terms do not only explain the appearance of secondary
$Q_2^+$, $A_X$ and $A_R$ modes but are also the key to understand the
different spin-orbital orders appearing in titanates. Although the two
types of antipolar distortions ($A_X$ -- X$_5^-$ irreps -- and $A_R$
-- R$_4^-$ irreps) have a distinct symmetry, their product belongs to
the irreducible representation of the JT motion ($X_5^- \cdot R_4^-$ =
$M_3^+$) as inferred by the fourth term of
Eq.~\ref{e:freeenergy}. Consequently, even in absence of significant
pristine $Q_2^+$ distortion, their joint appearance corresponds to an
effective Jahn-Teller motion that will drive orbital and spin
orders. As the tolerance factor decreases from large to small R
cations, the oxygen rotations ($\Phi_{xy}^-$ and $\Phi_z^+$ modes)
progressively increase, yielding larger anti-polar motions ($A_X$ and
$A_R$ modes) resulting from the first and third terms of
Equation~\ref{e:freeenergy} (see also Table~\ref{t:sym-exp}).  Then,
due to the effective JT character of these combined anti-polar
motions, their amplification produce an orbital ordering, comparable
to the one that would be produce by a proper $Q_ 2^+$ mode and which
is able to switch the magnetic ordering from G-type AFM to
FM. Remarkably, as confirmed by Table~\ref{t:sym-exp}, the
amplification of the effective JT mode is automatically accompanied by
a reduction of the proper JT mode, indicating a competition between
these two motions. The oxygen motions force together the appearance
and direction of the $Q_2^+$, $A_X$ and $A_R$ modes through the first
three terms in Equation~\ref{e:freeenergy}. Then, the fourth term
teaches us if the joint presence of these modes is by itself
energetically favorable, the $d$ coefficient of
Equation~\ref{e:freeenergy} is always found positive meaning that the
$A_X \cdot A_R \cdot Q_2^+$ trilinear term corresponds to an energy
penalty whose contribution has to be minimized. This progressive
disappearance of the $Q_2^+$ distortion as the tolerance factor
decreases further confirms its negligible character, and {\em de
  facto} the importance of the effective JT motions, on the
spin-orbital properties of rare-earth titanates.

We can check the role of the effective Jahn-Teller motion in our
calculations by analyzing in details the electronic and magnetic
properties of YTiO$_3$ and LaTiO$_3$. We observe that both YTiO$_3$
and LaTiO$_3$ are insulators in our simulations with band gaps of 1.04
and 0.94 eV respectively, compatible with experimental reports on
differentq
titanates~\cite{Tokura-bandgapRVO3,okimoto1995optical,loa2007crystal,grisolia2014structural,Gap-GdTiO3}. Our
computed magnetic moments on Ti$^{3+}$ are evaluated around 0.94
$\mu_B$ and 0.84 $\mu_B$ for YTiO$_3$ and LaTiO$_3$ respectively and
agree with experiments although the latter value is slightly
overestimated \cite{goral1983magnetic,eitel1986high}. However, all Ti
sites are occupied by only one electron.

\begin{figure}
\begin{center}
\resizebox{7.0cm}{!}{\includegraphics{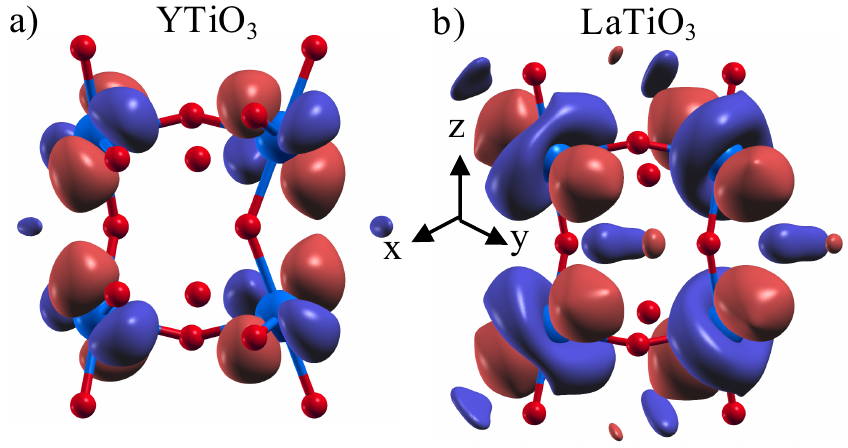}}
\end{center}
\caption{Orbital-orderings developed by out optimized ground state of
  YTiO$_3$ (a) and LaTiO$_3$ (b).}
\label{f:orbitalorderings}
\end{figure}

In the search of the localization of this single Ti-$d$ electron, we
built the Maxi-Localized Wannier Functions (MLWFs) for the ground
state of both materials using the Wannier90
software~\cite{Wannier90-1,Wannier90-2,Wannier90-3}. Firstly, we have
followed the strategy discussed in
reference~\cite{Nickelates-Julien}. We have initially projected the
Kohn-Sham states onto three generic $t_{2g}$ orbitals per Ti sites in
order to extract the initial gauge matrix for the localization
procedure. The latter is then restricted to the occupied manifold in
order to extract only occupied levels, albeit it is reduced to bands
with dominant O and Ti characters. The optimization renders only one
$t_{2g}$-like MLWFs per Ti site, and other MLWFs results in O-$p$
states in the vicinity of Ti sites. It further confirms the occupancy
of Ti $3d$ states by a single electron, whose localization renders the
orbital-orderings depicted in Figure~\ref{f:orbitalorderings}. These
orbital-orderings are very similar to those reported on the basis of
Dynamical Mean Field Theory calculations~\cite{Pavarini-PRL-RTO} as
well as reference~\onlinecite{PhysRevB.58.6831} for YTiO$_3$ and
references~\onlinecite{LTO-nondegenerate,Mochizuki-PRL-RTO} for
LaTiO$_3$. However, the shape of the resulting $t_{2g}$ orbital on
each Ti site can not be explained by a single electron lying in a
particular $t_{2g}$ orbital~\cite{mochizuki2001origin}. We can deduce
the different contributions of the $t_{2g}$ levels on the orbital
ordering by using different set of bands for the localization
procedure. To that end, we considered a total of 12 bands
corresponding to dominantly Ti $t_{2g}$ contributions located around
the Fermi level $E_F$, {\em i.e.} four bands below $E_F$, eight bands
above $E_F$. We then integrate the density of states projected on the
new $t_{2g}$-like WFs up to the Fermi level in order to extract their
contribution to the orbital-ordering~\cite{noteDOS-WFs}. We end with
very different contributions of the $t_{2g}$ states to the resulting
orbital-ordering:
\begin{eqnarray}
\left| \Psi_{YTiO_{3}} \right>  &\propto&  0.686 \left| d_{xy} \right> + 0.728 (\alpha \left|d_{xz}\right> + \beta \left|d_{yz}\right>)\\
\left| \Psi_{LaTiO_{3}} \right> &\propto&  0.565 \left| d_{xy} \right> + 0.825 (\alpha \left|d_{xz}\right> + \beta \left|d_{yz}\right>)
\end{eqnarray}
where $\alpha$ and $\beta$ are coefficients describing the
contribution of both $d_{xz}$ and $d_{yz}$ locally on each Ti sites
($\alpha^2+\beta^2$=1). It then appears that going from R=Y to R=La,
the orbital-ordering changes from a rather well balanced combination
of the $d_{xy}$ and ($d_{xz}$+$d_{yz}$) orbitals to a dominant
($d_{xz}$+$d_{yz}$)
character~\cite{mochizuki2001origin,Pavarini-PRL-RTO}.

\begin{figure}
\begin{center}
\resizebox{7.0cm}{!}{\includegraphics{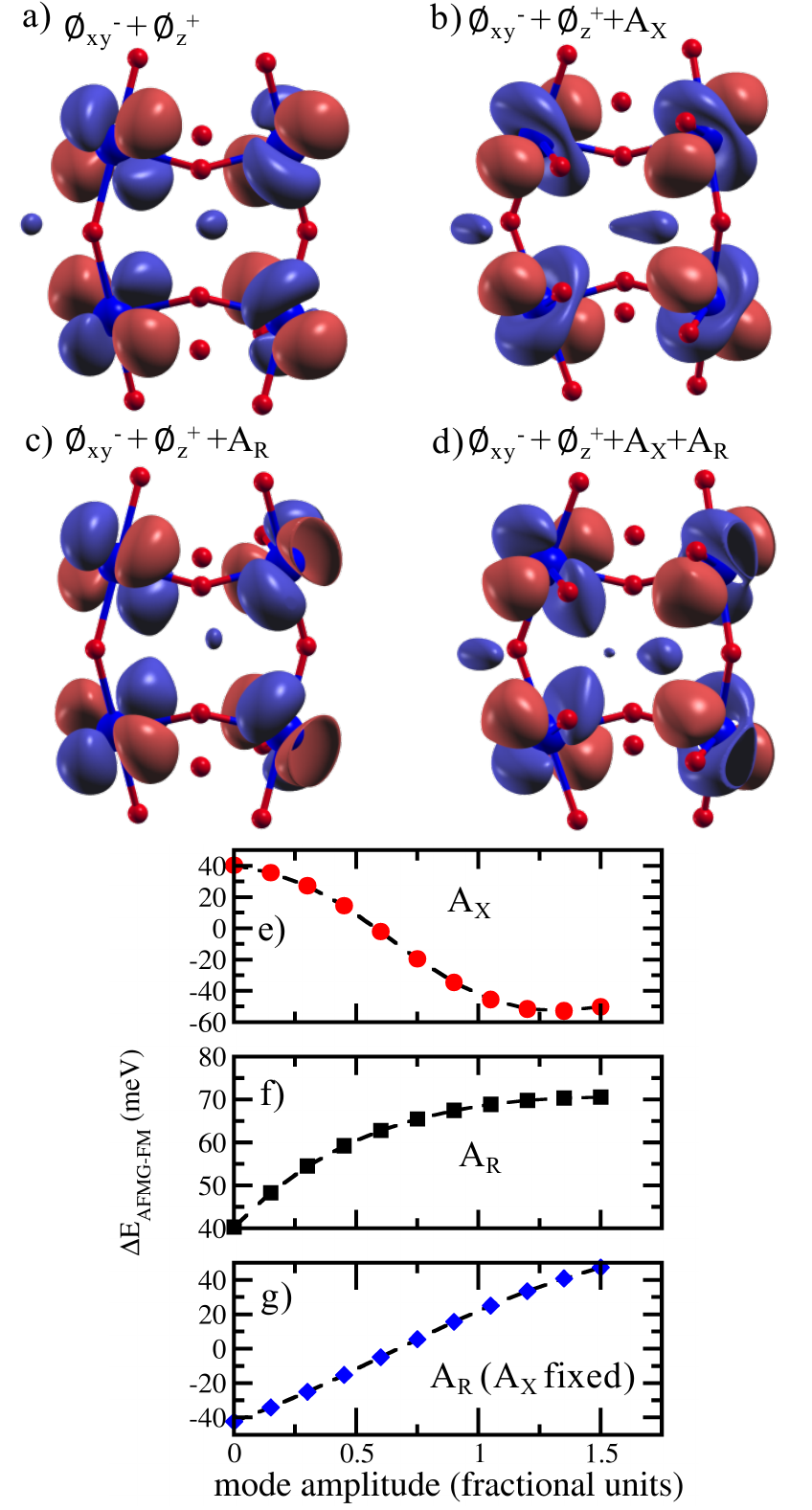}}
\end{center}
\caption{Influence of the rare-earth motions on the orbital and spin
  degrees of freedom. a) Orbital-ordering appearing with only the two
  oxygen cage rotations. b) Orbital-ordering obtained by freezing the
  antipolar X$_5^-$ motion with the two rotations. c) Orbital-ordering
  obtained by freezing the antipolar R$_4^-$ motion with the two
  rotations. d) Orbital-ordering obtained by freezing the antipolar
  R$_4^-$ motion with the two rotations and the antipolar $X_5^-$
  mode. e) and f) Energy difference between the G-AFM and FM solutions
  when adding either the antipolar X$_5^-$ (e), R$_4^-$ (f) antipolar
  motions to the system with the rotations. (g) Energy difference
  between the G-AFM and FM solutions when adding the antipolar $R_4^-$
  mode to the system with rotations and the $X_5^-$ mode.}
\label{f:influence}
\end{figure}

To gain insights on whether the contributions of the antipolar
distortions and the existence of the effective JT mode drive the
varying spin-orbital orders in titanates, we can track the evolution
of the orbital-ordering upon condensing different lattice modes
appearing in YTiO$_3$ in an ideal cubic phase having the ground state
volume. Starting from a structure with the two oxygen cage rotations,
we obtain an orbital ordering resembling that of YTiO$_3$ (see
Figure~\ref{f:influence}.a) with a minimal $d_{xy}$ orbital
contribution ($\left| \Psi\right> \propto 0.360 \left| d_{xy} \right>
+ 0.933 (\alpha \left|d_{xz}\right> + \beta \left|d_{yz}\right>$). On
the one hand, adding the $A_X$ antipolar mode to the two rotations
strongly suppresses the $d_{xy}$ character of the orbital-ordering,
the latter almost vanishes ($\left| \Psi\right> \propto 0.224 \left|
d_{xy} \right> + 0.975 (\alpha \left|d_{xz}\right> + \beta
\left|d_{yz}\right>$), and therefore the orbital-order is very similar
to that of LaTiO$_3$ (see Figure~\ref{f:influence}.b). On the other
hand, adding the $A_R$ antipolar motion to the rotations completely
switches the weight of the $d_{xy}$ and $d_{xz}$/$d_{yz}$ character
($\left| \Psi\right> \propto 0.706 \left| d_{xy} \right> + 0.709
(\alpha \left|d_{xz}\right> + \beta \left|d_{yz}\right>$) with an
orbital ordering now resembling that of YTiO$_3$ (see
Figure~\ref{f:influence}.c). Looking at the energy difference between
FM and G-AFM solutions when condensing independently the two antipolar
motions to the rotations, we observe that the $A_R$ mode favors a FM
ordering while the $A_X$ mode strongly enhances the stability of G-AFM
solution (see Figures~\ref{f:influence}.e and~\ref{f:influence}.f).

Therefore, starting from a structure with oxygen cage rotations and
the sole $A_X$ antipolar mode, all titanates should exhibit an
antiferromagnetic ordering of the Ti$^{3+}$ lattice (see
Figure~\ref{f:influence}.g). However, upon adding the $A_R$ lattice
distortion, the effective JT mode enters and is able to enhance the
$d_{xy}$ character of the orbital-ordering ($\left| \Psi\right>
\propto 0.419 \left| d_{xy} \right> + 0.908 (\alpha
\left|d_{xz}\right> + \beta \left|d_{yz}\right>$) and to stabilize the
FM ordering over the G-AFM solution (see Figures~\ref{f:influence}.d
and~\ref{f:influence}.g). We emphasize that the stability of the FM
order versus the G-AFM order is strongly enhanced in comparison to the
case of the sole condensation of $A_R$ with the rotations, further
proving the importance of the effective JT mode on the varying
spin-orbital properties of titanates. Finally, we observe that the
tetragonality $c/\sqrt{2}a$ of the unit, scaling with the oxygen cage
rotations amplitude, further increases the $d_{xy}$ contribution to
the orbital-order stabilizing the FM solution. It is worth to
emphasize that we do observe that the $Q_2^+$ mode has a rather
marginal effect leaving the weight of the three $t_{2g}$ orbitals on
the orbital-ordering unchanged.

In conclusion, our {\em first-principles} simulations highlight the
subtle interplay between structural, orbital and spin degrees of
freedom in rare earth titanates. Most notably, we have shown that, in
spite of the absence of sizable proper JT distortions, the oxygen
rotations inherent to the $Pbnm$ phase drive the appearance of two
distinct antipolar rare-earth motions, which together correspond to an
effective JT distortion. As the tolerance factor decreases, these
antipolar motions increases together with the oxygen rotations and
promote an orbital ordering similar to that which would be produced by
proper JT motions and favors a FM spin order explaining the change of
ground state from La to Y.  The anti-polar motions being generic to
any perovskite adopting a $Pbnm$ structure, this study demonstrates
that it is possible to achieve orbital orders and insulating phases in
these materials irrespectively of proper Jahn-Teller distortions.

\acknowledgments This work has been supported by the European Research
Consolidator (ERC) grant MINT under the contract
\#615759. Calculations took advantages of the Occigen machines through
the DARI project EPOC \#A0020910084 and of the DECI resource FIONN in
Ireland at ICHEC through the PRACE project FiPSCO. PhG acknowledges
support from the ARC project AIMED and F.R.S-FNRS PDR project
HiT4FiT. J. Varignon is very grateful to J. \'I\~{n}iguez for his
initiation to Wannier functions. Authors acknowledge J. Santamaria for
fruitful discussions.


\begin{thebibliography}{70}
\bibitem{NotePotentiels} We note that it is nearly impossible to
  converge the electronic wavefunction for single lattice distortions
  using the G-AFM order.
\bibitem{NoteInstabilite} We note here that when using the LaTiO$_3$
  ground state volume, the Jahn-Teller $Q_2^+$ motion develops an
  electronic instability for both YTiO$_3$ and LaTiO$_3$. Indeed at
  zero amplitude of the $Q_2^+$ mode, an energy gain of 1.90 and 0.57
  meV is achieved for YTiO$_3$ and LaTiO$_3$ respectively when the
  symmetry of the wavefunction is removed in the calculation, meaning
  that the electronic wavefunction does naturally want to distort to
  get rid off the electronic degeneracy.
\bibitem{noteDOS-WFs} The cartesian axes have been aligned along Ti-O
  directions in order to extract as much as possible pure $t_{2g}$
  orbitals.
\end{thebibliography}
\end{document}